\documentclass[twocolumn]{aastex631}

\usepackage{amsmath}
\usepackage{bm}
\usepackage{etoolbox}
\usepackage{booktabs}
\usepackage[stretch=30]{microtype}
\usepackage{graphicx}
\usepackage{hyperref}
\usepackage{fontawesome5}

\makeatletter
\patchcmd{\titleblock@produce}
  {\expandafter\produce@RRAP\expandafter{\centerline{\@submitted\hbox to 20pt{\hfill}}\vskip12pt}}
  {\noindent\makebox[\linewidth]{\@submitted}\par\vskip12pt}
  {}{\PackageError{EuclidFM}{Could not adjust the AASTeX submission line}{}}
\makeatother

\newcommand{\asinh}{\operatorname{asinh}}
\newcommand{\dmodel}{d_{\rm model}}

\begin{document}

\title{EuclidFM: Foundation Model for Compact Photometric Survey Representation}

\author{Reyhaneh Vojoudi Salmani}
\affiliation{School of Physics, Institute for Research in Fundamental Sciences (IPM), P.O. Box 19395-5531, Tehran, Iran}

\author{Mahsa Berahman}
\affiliation{Department of Physics, Shahid Beheshti University, P.O. Box 1983963113, Tehran, Iran}

\author{Saeed Ansarifard}
\affiliation{School of Physics, Institute for Research in Fundamental Sciences (IPM), P.O. Box 19395-5531, Tehran, Iran}
\correspondingauthor{Saeed Ansarifard}
\email{ansarifard@ipm.ir}


\author{Michael M. Fausnaugh}
\affiliation{Department of Physics and Astronomy, Texas Tech University, Lubbock, TX 79409-1051, USA}

\author{Seyed Mohammad Sadegh Movahed}
\affiliation{Department of Physics, Shahid Beheshti University, P.O. Box 1983963113, Tehran, Iran}

\shorttitle{EuclidFM}
\shortauthors{Vojoudi Salmani et al.}

\begin{abstract}
We present \textsc{EuclidFM}, a compact self-supervised model that learns object-level representations from heterogeneous catalogue photometry in Euclid Quick Data Release 1. Using 90,258 sources across three Euclid Deep Fields, a masked-autoencoder Transformer reconstructs withheld fluxes and uncertainties from available measurements, producing a 128-dimensional embedding per source. The embeddings encode morphological and photometric information: sources with high point-like probability have, on average, 12.4 similarly classified neighbours among their 15 nearest neighbours, compared with 4.3 under field- and magnitude-stratified randomization. Linear models trained on frozen embeddings predict VIS magnitudes with correlations of 0.954–0.980 across three held-out fields. For sources with VIS aperture-flux uncertainties up to $1\,\mu\mathrm{Jy}$, incorporating embeddings reduces the root-mean-square error in predicted log uncertainty by at least 24\% relative to a magnitude-only model. A classifier trained on spectroscopically confirmed sources identifies 1,150 quasar candidates, but only 21 of the 156 candidates with accepted external counterparts show AGN evidence, including 12 spectroscopic quasars. The classifier's transferability to the wider photometric catalogue therefore remains uncertain, and its purity and completeness require further evaluation. Independently of classification, EuclidFM provides a common representation for downstream survey analyses when photometric coverage differs between sources. Code and data are available on~\href{https://github.com/AI-Driven-HEP/EuclidFM}{\faGithub}.
\end{abstract}


\section{Introduction}
\textit{Euclid} maps the large-scale structure of the Universe to
investigate dark matter and dark energy \citep{EuclidMellier2025}.
Its first Quick Data Release (Q1) contains about 30 million sources
across the three Euclid Deep Fields (hereafter, fields)
\citep{EuclidQ1Overview2026},
and DR1-Foundation extends the catalogue to more than one billion
sources \citep{EuclidDR1Overview2026}. The MERge (MER) catalogues
combine Euclid and ground-based photometry
\citep{EuclidMER2026,EuclidDR1MER2026}, providing broad wavelength
coverage but a heterogeneous set of measurements for individual sources.

This heterogeneity reflects differences in survey coverage, depth,
and measurement types. For example, northern and southern
Wide Survey $g$-band imaging comes primarily from HSC and DECam,
respectively \citep{EuclidDR1External2026}. Spatial variations in
depth and data quality further affect the available photometry
\citep{EuclidDR1Masks2026}. Requiring complete measurements in a
fixed set of channels can therefore discard useful information
or exclude sources. Learning from the available measurements,
their reported uncertainties, and their provenance offers a way
to represent these catalogues without requiring uniform coverage.

\begin{figure}[t!]
\centering
\includegraphics[width=\columnwidth]{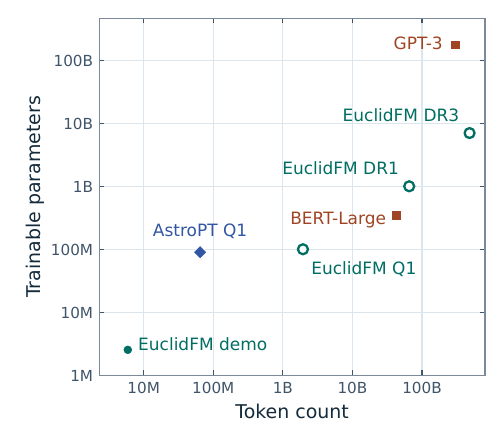}

\caption{Model sizes and token counts for EuclidFM and selected reference
models. EuclidFM includes $5.91\times10^6$ flux--uncertainty pairs
across all 90,258 sources. Open circles extrapolate the mean
measurements per source to full catalogues at illustrative model
sizes; DR3 assumes the DR1 source density over the full Wide
Survey area \citep{EuclidDR1Release2026,EuclidMellier2025}.
Astronomical dataset inventories and cumulative language-model
training counts are not directly comparable (see text).}

\label{fig:euclidfm-scale}
\end{figure}

Foundation models seek reusable representations through pretraining
and adaptation to downstream tasks \citep{Bommasani2021}.
In astronomy, AstroPT learns by predicting successive image patches
\citep{Smith2024AstroPT}; its Euclid Q1 application also incorporates
near-infrared images and broadband spectral energy distributions
\citep{Siudek2026}. AstroCLIP aligns separately pretrained image
and spectrum encoders through contrastive learning
\citep{Parker2024}, an approach subsequently adapted to matched
Euclid images and DESI spectra \citep{Ronceray2026}.
Related work addresses heterogeneous stellar measurements
\citep{LeungBovy2024} and joint modelling of imaging, spectra,
and scalar catalogue data \citep{Parker2025AION}.
These developments motivate a complementary question: how much
useful source information can a compact model retain using
heterogeneous Euclid catalogue~photometry~alone?

In this work, we present \textsc{EuclidFM}, a compact
self-supervised representation model and a step toward
catalogue-level foundation models. We use a working sample of
90,258 Euclid Q1 sources, with an 80:20 pretraining--validation
split. Each photometric measurement is identified by its survey, band, and measurement type, with the corresponding flux--uncertainty pair represented as a token.
A masked autoencoder \citep{He2022} reconstructs deliberately
withheld measurements from the remaining tokens; naturally
unavailable measurements are excluded from attention and the
reconstruction loss. The encoder uses bidirectional context,
as in masked prediction methods such as BERT
\citep{Devlin2019BERT}, rather than the preceding-token context
used by AstroPT. Reported uncertainties enter as measured
features and reconstruction targets.

Figure~\ref{fig:euclidfm-scale} places this demonstration in the
context of larger models and illustrative catalogue expansions.
The EuclidFM and AstroPT points count available photometric pairs
and VIS patches, respectively, once across their datasets.
BERT-Large counts cumulative input positions, including padding,
and GPT-3 counts cumulative text tokens processed during training
\citep{Devlin2019BERT,Brown2020GPT3}. These different conventions
provide scale context, not a comparison of computational efficiency
or a demonstrated scaling relation.

The pretrained encoder yields a 128-dimensional vector from a
learned summary (CLS) token for each source. We test associations
with MER morphology in the original embedding space and examine
whether VIS brightness can be recovered across Euclid Deep Fields (hereafter,
fields). We also
visualise how survey coverage shapes the representations.
As an illustrative downstream task, we train a small
quasar--star--galaxy classifier on frozen embeddings of a curated
labelled reference sample. We apply the calibrated classifier to
the working catalogue and examine the resulting candidates using
Gaia and other external catalogues. This tests the information
retained by the representations while distinguishing candidate
classification from spectroscopic confirmation.

Section~\ref{sec:data_methods} describes the 90,258 source working catalogue, EuclidFM pretraining, and the source classifier.
Section~\ref{sec:results} examines the learned embeddings and presents the stellar and quasar candidates and their external matches.
Section~\ref{sec:discussion} discusses the implications and limitations of these results.

\section{Data and Methods}
\label{sec:data_methods}

\begin{figure*}[!t]
\centering
\includegraphics[width=\textwidth]{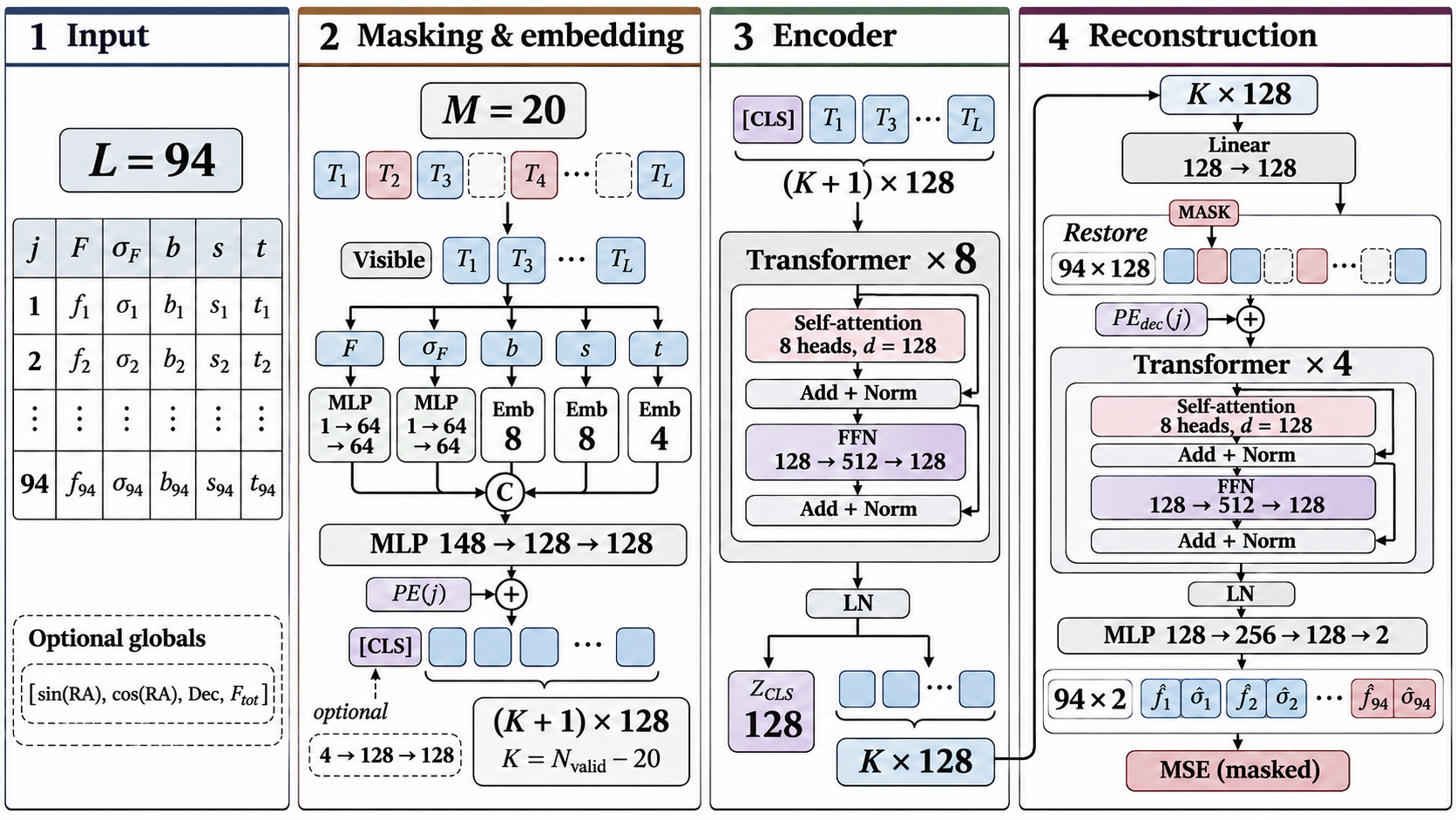}
\caption{EuclidFM masked-autoencoder architecture. Only unmasked available
measurements enter the encoder; continuous flux/error projections are fused
with categorical metadata and original-index positional embeddings. A
\texttt{[CLS]} token provides the object representation. Measurement
embeddings are restored to the full channel order for four-layer decoder
reconstruction, and the loss is evaluated only at masked available
positions. Blue, pink, and dashed gray tokens indicate visible, withheld,
and unavailable measurements, respectively. C denotes concatenation and
PE denotes learned positional embeddings. The optional global-feature
branch is disabled in the reported configuration.}
\label{fig:architecture}
\end{figure*}

\subsection{Catalogue and Photometric Channels}

The working catalogue is drawn from an initial random sample of
100,000 Q1 objects, allocated approximately equally among EDF-North,
EDF-South, and EDF-Fornax. Quality filtering and additional exclusions
leave the 90,258 sources introduced above; the selection is detailed
in Appendix~\ref{sec:appendix_sample}. The photometry includes
Euclid VIS and Near Infrared Spectrometer and Photometer (NISP)
measurements, together with DECam, HSC, Pan-STARRS, and MegaCam
measurements from the Q1 catalogue products.

Each combination of survey, band, and measurement type defines a
\emph{channel}. The selected photometry spans 94 channels, including
different aperture sizes and model-fitting measurements of the same
band. These measurements can share underlying images and need not
be independent. A channel is available for a source when both its
flux and reported uncertainty are finite. Under this criterion,
sources have 29--70 available channels, with a median of 64.

We apply the inverse hyperbolic sine transformation separately to
fluxes and uncertainties to compress their dynamic range while
retaining zero and negative fluxes. The transformed values are
combined with survey, band, and measurement-type embeddings.
Learned positional embeddings preserve each channel's index in a
fixed wavelength-ordered layout. The channel inventory and input
transformation are specified in
Appendices~\ref{sec:appendix_channels} and
\ref{sec:appendix_flux_transform}.

\subsection{Pretraining and Embedding Extraction}
\label{sec:pretraining_extraction}

Figure~\ref{fig:architecture} shows the data flow from catalogue
measurements to the source embedding. During pretraining, we randomly
withhold 20 available measurements per source at each presentation.
Each remaining flux--uncertainty pair is embedded as a 128-component
token using its metadata and original channel position. An eight-layer
Transformer encoder \citep{Vaswani2017} processes those tokens together
with a learned summary token (\texttt{[CLS]}) at the start of the
sequence. Only the measurement outputs go to the four-layer decoder,
which restores the original channel order and predicts the withheld
fluxes and uncertainties. The mean squared reconstruction loss is
evaluated only on the withheld measurements. The reconstruction mask
is resampled, whereas the availability mask is fixed by the catalogue.
No source-class labels, sky coordinates, or explicit morphology
diagnostics enter pretraining.

For source $i$, let $K_i$ be the number of measurement tokens passed
to the encoder and $\bm{X}_i$ the input sequence, including
\texttt{[CLS]}. Ignoring padding, the encoder returns
\begin{equation}
\bm{H}_i = \operatorname{Enc}_{\theta}(\bm{X}_i)
\in \mathbb{R}^{(K_i+1)\times128}.
\end{equation}
The first row (index zero) is the contextualised \texttt{[CLS]}
output, $\bm{z}^{\rm CLS}_i=\bm{H}_i[0,:]\in\mathbb{R}^{128}$.
The remaining $K_i$ rows correspond to the measurement tokens.
For a batch of $B$ sources padded to $K_{\max}=\max_i K_i$
measurement tokens, the encoder output has shape
$B\times(K_{\max}+1)\times128$, and taking the first row of each
sequence gives a $B\times128$ array of CLS vectors.

We use the encoder weights from the training epoch with the lowest
validation reconstruction loss. To extract embeddings, we keep those
weights fixed and use all available measurements, so $K_i$ is the
number of available channels for source $i$. The representation
analyses and classifier use only the 128-component CLS vector per
source, not the measurement outputs or the decoder. Architecture and
training details are given in Appendix~\ref{sec:appendix_pretraining}.

\subsection{Reference Sample}

The labelled reference set, termed the \emph{Platinum Sample},
contains 9,675 sources: 2,675 quasars, 3,500 stars, and 3,500
galaxies. Quasars are drawn from the spectroscopic
\emph{Golden Sample} of \citet{Fu2026}, while stellar and galaxy
labels are supported by external astrometric and morphological
information. Sample construction is described in
Appendix~\ref{sec:platinum}.

\subsection{Classifier}
\label{sec:classifier}
The supervised classifier is a multilayer perceptron (MLP) operating on the 128-dimensional CLS representations, $\bm{z}^{\rm CLS}$, extracted by the frozen pretrained EuclidFM encoder. The classifier consists of two fully connected hidden layers with 64 and 32 neurons, respectively, each followed by a ReLU activation, and a final linear layer with three output units corresponding to QSOs, stars, and galaxies. The network is trained using the cross-entropy loss and the Adam optimizer with a learning rate of $10^{-3}$. Only the classifier parameters are optimized, while the pretrained EuclidFM encoder remains frozen. We use a class-stratified 70:15:15 split for
classifier training, validation, and testing. The classifier achieves an overall accuracy of 97.3\% on test set. Table~\ref{tab:platinum-classification} summarizes the classification performance using the class with the highest predicted probability (argmax) as the decision rule.

\begin{table}[!htbp]
\centering
\caption{Classification performance on the Platinum Sample using a
supervised classifier trained on frozen EuclidFM representations.
Precision, recall, and F1-score are reported for each class.}
\label{tab:platinum-classification}
\noindent\begin{tabular}{@{}lrrr@{}}
\toprule
Class & Precision & Recall & F1-score \\
\midrule
QSO    & 0.953 & 0.950 & 0.951  \\
Star   & 0.991 & 0.994 & 0.992  \\
Galaxy & 0.971 & 0.970 & 0.970 \\
\midrule
Accuracy      &        &        & 0.973  \\
Macro average & 0.972  & 0.971  & 0.971  \\
Weighted avg. & 0.973  & 0.973  & 0.973  \\
\bottomrule
\end{tabular}
\end{table}

Having evaluated the classifier on the Platinum Sample, we apply it to the working catalog to assign labels to previously unlabeled objects. To obtain a high-confidence sample, we adopt class-specific probability thresholds instead of highest predicted probability, chosen on the validation set of the Platinum Sample to give a precision above 99\% for each class. The calibrated probabilities thresholds are 0.83 for QSOs, 0.96 for stars, and 0.95 for galaxies. A source receives its highest-probability class only when that probability meets the corresponding threshold; otherwise it remains unclassified. At these operating points the recalls are 84.8\%, 97.3\%, and 83.6\%, respectively; 89.7\% of the validation sample receives a label, and the remaining 10.3\% is left unlabeled. Among the unlabeled objects, galaxies make up the largest fraction, followed by QSOs and stars, consistent with the lower recall of galaxies and QSOs at these thresholds.

We apply the same thresholds to the catalog and use the resulting labels to study its class composition and the distribution of classes in the learned representation space. The quoted precision applies to the Platinum validation sample, which is class-balanced and brighter than most of the catalog. In the full catalog, where galaxies dominate and many sources are fainter than those in the Platinum Sample, we find evidence that the purity of the QSO labels are substantially lower (see \ref{sec:results}). We therefore report the Precision and Recall with the caveat that they are only applicable to the validation sample.



\section{Learned Representations and Source Classification}
\label{sec:results}

We visualise local structure in the frozen CLS vectors of the 90,258
sources with one two-dimensional t-SNE projection
\citep{vanDerMaaten2008}. We retain 50 unwhitened principal components (99.72\% of the
CLS variance) and use a perplexity of 30 and a random seed of 42.
The catalogue properties and classifier outputs are overlaid after
projection. Distances used in the quantitative tests below are
Euclidean distances in the original 128-dimensional CLS space.

\subsection{Apparent Morphology and Deblending}

\begin{figure*}[!tp]
\centering
\includegraphics[width=0.73\textwidth]{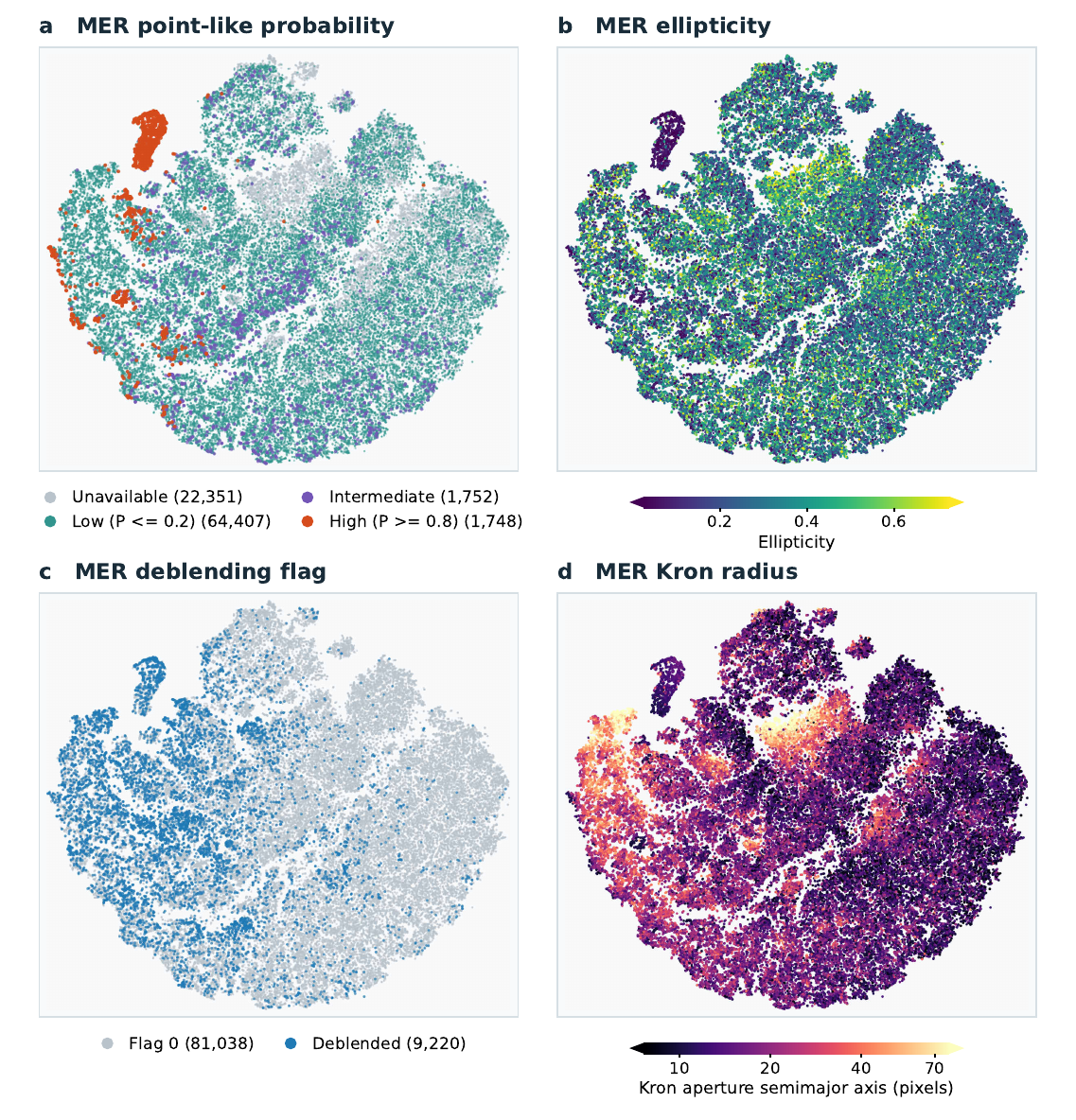}
\caption{MER morphology on the t-SNE projection of the
90,258-source working catalogue. (a) Point-like probability
$P_{\rm point}$, grouped into low ($\leq 0.2$), intermediate, high
($\geq 0.8$), and unavailable values. (b) Ellipticity.
(c) MER deblending flag. (d) Catalogue Kron radius in pixels, shown
on a logarithmic colour scale. The continuous scales span the 1st to
99th percentiles.}
\label{fig:tsne-mer-morphology}
\end{figure*}

Figure~\ref{fig:tsne-mer-morphology} shows a compact locus of
sources with high MER point-like probability and low ellipticity.
Although the encoder receives no explicit morphology diagnostics,
aperture and model-fit fluxes can carry information about apparent
structure. The MER point-like probability itself depends on
photometric measurements of brightness and concentration
\citep{EuclidMER2026}. We test whether the grouping persists in CLS
space among 66,106 sources with measured VIS magnitudes and either
$P_{\rm point}\leq 0.2$ or $P_{\rm point}\geq 0.8$. To account for
differences across fields and in brightness, we use a random baseline
that keeps the fraction of high-probability sources fixed in each
0.25-mag VIS interval within each field
(Appendix~\ref{sec:appendix_cls_tests_2}). For each high-probability
source, an average of 12.4 of its 15 nearest CLS neighbours also have
high probability, compared with 4.3 expected under this baseline.
The excess indicates that the embeddings retain information associated
with apparent morphology beyond the differences captured by field and
brightness alone.

Above the centre of the projection, a patch of high ellipticity
coincides with large Kron radii and many unavailable point-like
probabilities. Large Kron radii also occur elsewhere, while flagged
deblends are concentrated on the left. Among 88,576 sources with
measured VIS magnitudes, those in the largest Kron-radius decile
have on average 2.85 flagged deblends among their 15 nearest CLS
neighbours, compared with 1.20 for the rest of the sample. The greater number of flagged neighbours around large-Kron-radius sources can be explained by two factors: large-Kron sources tend to lie near other large-Kron sources in CLS space, and large-Kron sources are more frequently flagged as deblended (Appendix~\ref{sec:appendix_cls_tests_2}).

\subsection{VIS Brightness in CLS Space}

\begin{figure*}[!tp]
\centering
\includegraphics[width=0.73\textwidth]{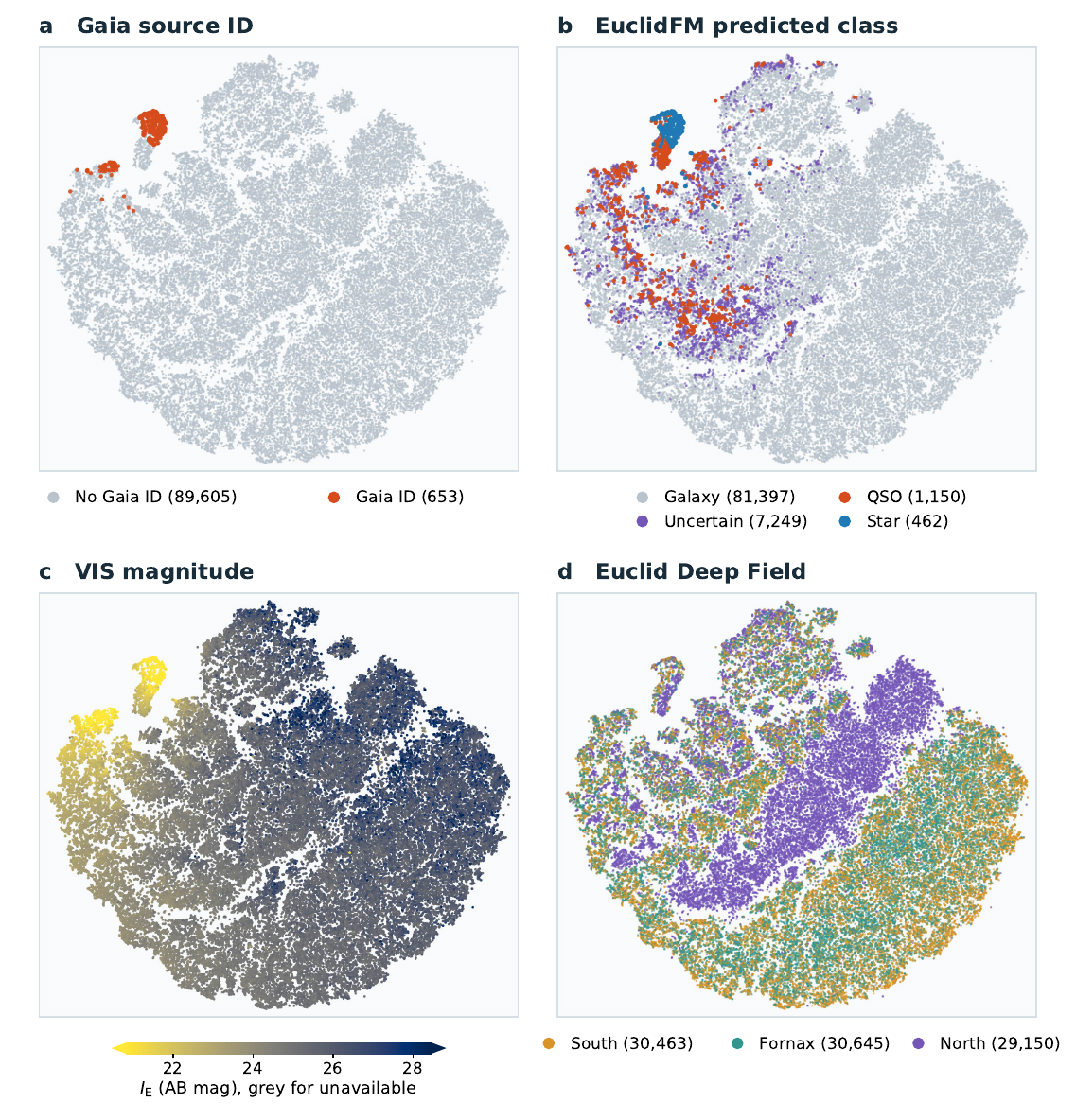}
\caption{Additional overlays on the t-SNE projection in
Figure~\ref{fig:tsne-mer-morphology}. (a) Presence of a published
Gaia DR3 source ID in MER. (b) EuclidFM assignments using the
class-specific probability thresholds of classifier.
Uncertain denotes sources whose highest-probability class fails its
threshold. (c) VIS $I_{\rm E}$ AB magnitude, with grey for unavailable
measurements and a colour scale spanning the 1st to 99th percentiles.
(d) Euclid Deep Field membership.}
\label{fig:tsne-class-field}
\end{figure*}

\begin{figure*}[!tp]
\centering
\includegraphics[width=\columnwidth]{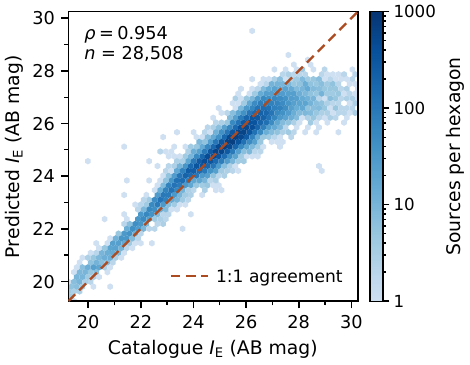}
\caption{Predicted versus catalogue VIS $I_{\rm E}$ AB magnitude in
EDF-North. A least-squares model fitted to the frozen CLS vectors in
EDF-Fornax and EDF-South predicts the North magnitudes. Hexagons are
coloured by source count on a logarithmic scale, and the dashed line
marks one-to-one agreement. The axes show the central magnitude range.
The Spearman rank correlation $\rho=0.954$ is evaluated over all
28,508 held-out sources.}
\label{fig:vis-magnitude-north}
\end{figure*}

The VIS overlay in panel (c) of Figure~\ref{fig:tsne-class-field}
shows a brightness gradient across the map. To quantify this trend,
we take each of the 88,576 sources with measured $I_{\rm E}$ a reference source. For each reference source, we randomly select eight pairs of sources from the same Euclid field, all with measured $I_{\rm E}$. The two sources in each pair must be different from each other and from the reference source. In 76.4\% of comparisons, the member closer to the
reference in CLS space is also closer to it in VIS magnitude. This
agreement exceeds the 49.76\%--50.25\% range from the within-field
random baseline (Appendix~\ref{sec:appendix_cls_tests_mid}), showing that
sources closer in CLS space tend to be more similar in VIS brightness
even within the same field.
A linear model fitted to predict $I_{\rm E}$ from CLS vectors in
two fields and evaluated in the third gives Spearman rank
correlations of $0.954$--$0.980$ across the three held-out fields (Figure~\ref{fig:vis-magnitude-north}).
This test measures how well the embeddings retain brightness
information present in the input photometry.

\subsection{VIS Measurement Quality in CLS Space}
\label{sec:vis_quality}

\begin{figure*}[!htbp]
\centering
\includegraphics[width=0.83\textwidth]{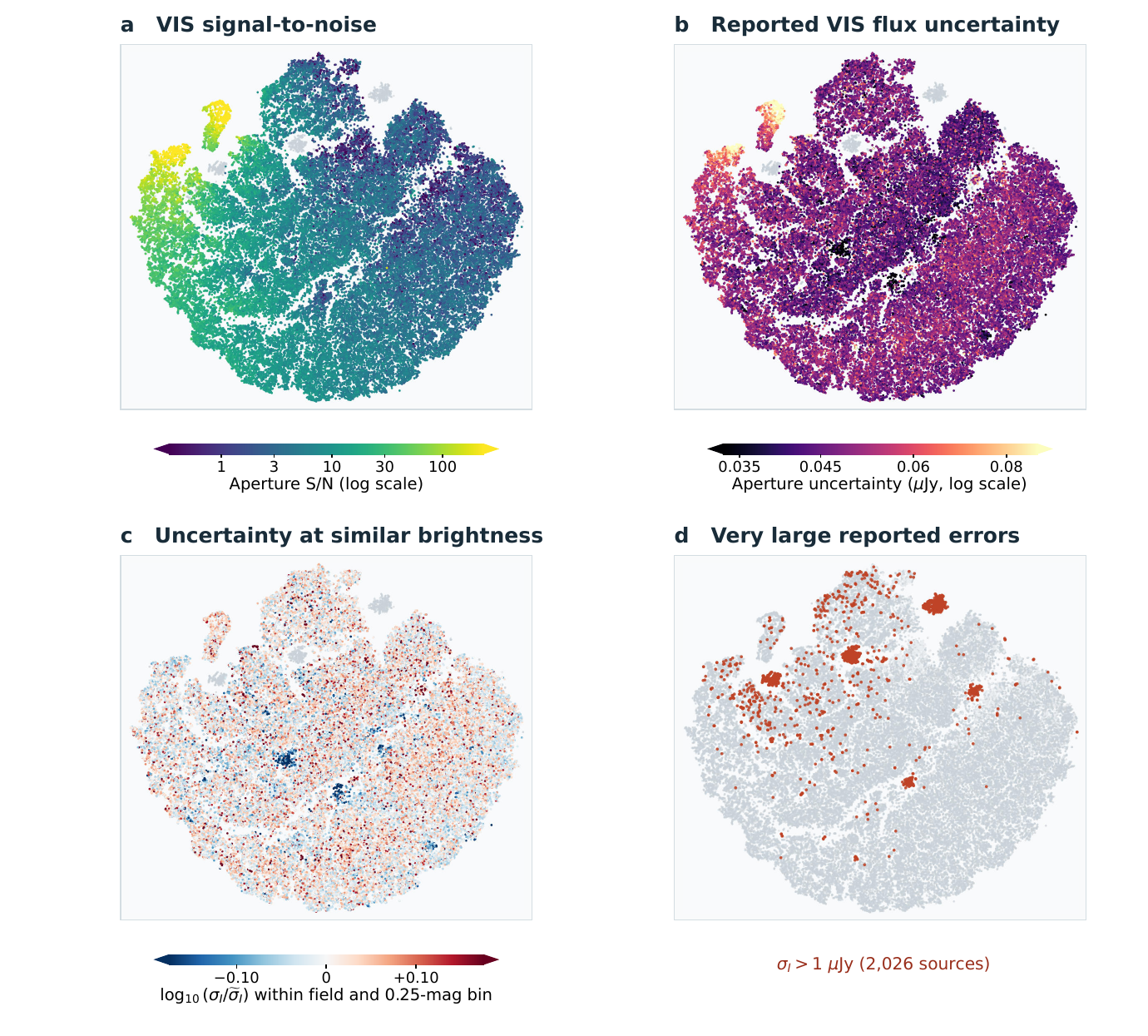}
\caption{VIS measurement quality on a t-SNE projection recomputed
for all 90,258 CLS vectors with the settings in Section~\ref{sec:results}.
(a) Aperture S/N. (b) Reported aperture flux uncertainty.
(c) Log uncertainty relative to the median in the same field and
0.25-mag VIS bin. Panels (a)--(c) show the 86,550 sources with positive
finite flux and uncertainty and $\sigma_I\leq1\,\mu\mathrm{Jy}$ in colour,
with the remaining sources in grey. Panel (d) highlights the 2,026
sources with $\sigma_I>1\,\mu\mathrm{Jy}$ and positive finite flux.}
\label{fig:vis-quality-overlays}
\end{figure*}

High VIS signal-to-noise ratio (S/N) occupies the bright side of the
CLS projection (Figures~\ref{fig:tsne-class-field} and
\ref{fig:vis-quality-overlays}). Within a field, the source closer in
CLS space is also closer in S/N in 74.9\% of the pair comparisons.
Restricting the comparisons to sources of similar VIS magnitude
reduces agreement to 52.3\%, close to the 50.0\% random baseline
(Appendix~\ref{sec:appendix_cls_tests_mid}). S/N is the ratio of the
aperture flux to its reported uncertainty $\sigma_I$ and correlates
strongly with brightness (Spearman $\rho=0.940$ for $-I_{\rm E}$).
The drop in agreement from 74.9\% to 52.3\% suggests that the similarity in S/N among nearby sources in CLS space is largely explained by their similar VIS brightness. In contrast, the agreement between CLS proximity and reported flux uncertainty is already close to random, changing only from 52.7\% to 52.5\% when comparisons are restricted to sources of similar brightness. Consistent with this, sources with uncertainties above or below the median for their field and magnitude bin are more widely distributed across the CLS projection, rather than concentrated in distinct regions.

The reported uncertainties have a high-error tail: 2,026 sources have $\sigma_I>1\,\mu\mathrm{Jy}$, compared with a sample median of $0.047\,\mu\mathrm{Jy}$. The Euclid Q1 MER release notes document inflated aperture uncertainties in regions observed with few dithers.\footnote{See the \href{https://euclid.esac.esa.int/dr/q1/expsup/Ch15.S2.SS3.html}{Euclid Q1 MER release notes}. We have not verified whether this effect explains the extreme uncertainties of the individual sources in our sample.} Excluding the tail as a sensitivity check leaves the local agreement near the random baseline: 51.7\% for S/N and 51.6\% for reported uncertainty after matching in field and magnitude.

Weak local agreement does not mean that the embeddings contain no
information about uncertainty. In the restricted sample, a linear
model using the CLS coordinates yields held-out Spearman correlations
of 0.946--0.978 for S/N and 0.556--0.712 for reported uncertainty,
compared with 0.100--0.223 for uncertainty in the full sample. More
directly, adding the CLS coordinates to a magnitude-only model
reduces the held-out root-mean-square error in log uncertainty by
24\%--37\% across fields in the restricted sample
(Appendix~\ref{sec:appendix_cls_tests_last}). Thus the embeddings retain
information about the catalogue's reported uncertainties beyond
brightness alone, even though those uncertainties do not strongly
organise the local neighbourhoods. This is a test of information
retention, not an independent calibration of the measurement errors.

\subsection{Structure Across the Deep Fields}

The field overlay (panel (d) of Figure~\ref{fig:tsne-class-field})
shows EDF-North concentrated in a central band and EDF-South and
EDF-Fornax overlapping extensively in the lower part of the
projection. The two southern fields share substantial DECam
coverage, while EDF-North draws on a different optical survey mix
\citep{EuclidQ1Overview2026}. That difference could contribute to
the field structure. All three fields also overlap in parts of the
map, consistent with shared Euclid measurements or similar source
properties shaping some local neighbourhoods. Testing their
relative contributions would require comparing embeddings under
matched photometric coverage.

\subsection{Catalogue Classification and Stellar Candidates}

Applying the calibrated probability thresholds of 0.83 for quasars,
0.96 for stars, and 0.95 for galaxies yields 1,150 quasar candidates, 462
stellar candidates, and 81,397 galaxy assignments. Another 7,249
sources remain unclassified. Stellar candidates and sources with published Gaia
DR3 IDs concentrate in the upper-left point-like locus, while quasar
candidates occupy several neighbouring regions
(Figure~\ref{fig:tsne-class-field}, panels (a) and (b)).

Of the 462 stellar candidates, 362 (78.4\%) have Gaia DR3
counterparts linked by the MER source IDs \citep{GaiaDR3}.
We group the stellar classes of Gaia's Discrete Source Classifier
(DSC) and require the 0.80 maximum class probability cut
\citep{Creevey2023Apsis}. One source falls below the cut. From the remaining 361 sources, gaia assigns 359 as stellar class (99.4\%) and two to the quasar
class.

\begin{figure*}[!htbp]
\centering
\includegraphics[width=0.63\textwidth]{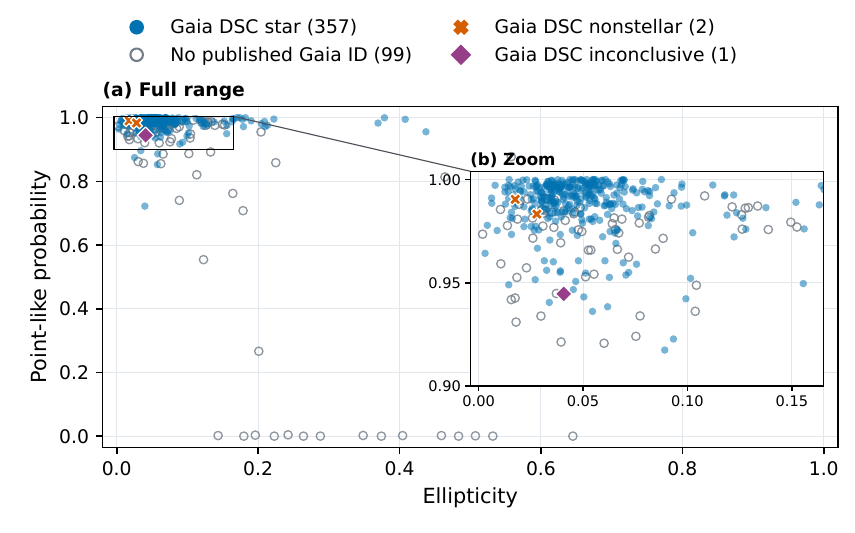}
\caption{MER point-like probability versus ellipticity
for 459 EuclidFM stellar candidates. Panel (a) shows the full plotted
range, and the inset (b) magnifies the high-probability,
low-ellipticity region. The symbols distinguish Gaia DSC stars (357),
Gaia DSC quasars (two, marked as nonstellar), one inconclusive Gaia
classification, and sources without a published Gaia ID (99).
Three candidates without a point-like probability are omitted.}
\label{fig:stellar-morphology-gaia}
\end{figure*}

Among the stellar candidates with measured morphology ond without Gaia ID counterpart, 15 have
$P_{\rm point}<0.01$ (Figure~\ref{fig:stellar-morphology-gaia}). All 15 carry Euclid PHZ galaxy labels and
lack published Gaia IDs. Their median ellipticity is 0.35, compared
with 0.047 for Gaia-classified stars in this sample, and nine carry
the MER detection-quality blending flag. These diagnostics flag
possible reason for the sources miscalssified as stars.

\subsection{Quasar Candidates and External Matches}

To assess the quasar selection, we combine positional associations
within $1^{\prime\prime}$ from external catalogues, counting each
Euclid source once. Among 869 sources with accepted counterparts,
38 have QSO or active galactic nucleus (AGN) evidence. These include
16 spectroscopic QSOs from DESI DR1, Milliquas, or SDSS DR20
\citep{DESIDR1_2026,Flesch2023Milliquas,SDSSDR20_2026}, along with
published identifications and candidate classifications. From these 38 objects, 21 are labeled as QSO in our classifiers. 10 objects are unknown, 5 as galaxy as 2 are labeld as stars. Table \ref{tab:external-classification} summarizes fraction labeled as quasars for each category. These fractions leave the purity and completeness of the full quasar candidate sample undetermined. However, it is clear that the classifier does not perform as well as on the Platinum validation sample. The poor performance suggest that the classifier cannot be transferred to our wider sample of 90,258 objects from the photometric catalog. The issue might be related to the small number of objects in the Platinum sample for training, limited magnitude range, criteria for inclusion in the golden sample, criteria for valid spectroscopic reductions, or sky coverage/blending  of the golden sample; all of these factors can make the golden sample unrepresentative of our EuclidFM sample, drawn from the photometric catalog.  Further work will investigate this bias and attempt to improve the classifier.

\begin{table}[t]
\centering
\caption{Comparison of EuclidFM classifier labels with external catalogue
classifications for 869 unique Euclid objects with accepted counterparts
within $1^{\prime\prime}$. Each object is counted once. The external Star
category combines stars identified by astrometry or spectroscopy with
likely stars from Gaia DSC. Unverified includes ambiguous cases.}
\label{tab:external-classification}
\footnotesize
\setlength{\tabcolsep}{1.5pt}
\begin{tabular}{@{}lrrrrr@{}}
\toprule
 & \multicolumn{4}{c}{EuclidFM classifier} & \\
\cmidrule(lr){2-5}
External   & Star & Galaxy & QSO & Unknown & Total \\
\midrule
Star       &  363 &      3 &  83 &     145 &   594 \\
Galaxy     &    0 &     40 &  39 &     107 &   186 \\
QSO/AGN    &    2 &      5 &  21 &      10 &    38 \\
Unverified &    1 &     12 &  13 &      25 &    51 \\
\midrule
Total      &  366 &     60 & 156 &     287 &   869 \\
\bottomrule
\end{tabular}
\end{table}


\section{Discussion and Conclusions}
\label{sec:discussion}

EuclidFM learns compact representations of heterogeneous catalogue photometry through masked reconstruction. A classifier trained on the frozen vectors selects 1,150 quasar candidates and 462 stellar candidates. To evaluate these classifications, we cross-match the candidates with external catalogues. To our knowledge, external counterparts were identified for fewer than 1\% of the 90,258 sources in the working catalogue, and not all of these counterparts have reliable classifications. Among the matched candidates, Gaia classifies 359 of the 361 stellar candidates that pass the Gaia probability cut as stars. Of the 156 quasar candidates with accepted external counterparts, 21 have QSO/AGN evidence, including 12 spectroscopic quasars. The purity and completeness of the full quasar selection remain to be established. Importantly, the Platinum sample used to train and evaluate the classifier is highly selective, whereas the working catalogue was constructed to resemble the broader Euclid catalogue as closely as possible. This difference in sample distributions may contribute to the discrepancy between classifier performance on the held-out Platinum test set and its results when applied to the working catalogue.

Masked reconstruction lets EuclidFM learn relationships among photometric channels from sources without class labels. The resulting CLS vector gives downstream models a common input even when the available measurements differ between sources. One advantage over training directly on catalogue photometry is that pretraining could reduce the number of labelled sources needed to reach a given level of performance. We defer a direct comparison between classification using CLS vectors and classification using the original catalogue measurements to future work.

The VIS tests show that the CLS vector preserves brightness across fields, while most of the agreement between proximity in CLS space and similarity in S/N disappears when sources are matched in field and magnitude. For the reported VIS aperture uncertainty, the linear readout yields correlations of $0.10$--$0.22$ when all sources with positive, finite VIS fluxes and uncertainties are included, and $0.56$--$0.71$ after excluding sources with $\sigma_I>1\,\mu\mathrm{Jy}$ (Section~\ref{sec:vis_quality}). In the restricted sample, adding the CLS coordinates to a magnitude-only model reduces the root-mean-square error in predicted log uncertainty by 24--37\% across held-out fields. The embedding therefore carries information about reported VIS uncertainties beyond what can be inferred from brightness alone, within the range covered by the restricted sample. Although EuclidFM uses all available photometric bands, we focus on VIS measurements to evaluate the CLS representations. Similar analyses of other photometric bands are left for future work.

We plan to extend pretraining to larger Euclid catalogues and broader coverage in sky position. The CLS vector can be combined with embeddings from image and spectroscopic models for the same sources. Downstream models can use this compact source vector or the full sequence of contextualised measurement-token embeddings. The latter include a separate representation for each available channel. These options could support tasks such as photometric redshift estimation, in addition to the classification demonstrated here.

\section*{Acknowledgments}
This work has made use of the Quick Release (Q1) data from the Euclid mission of the European Space Agency (ESA). This research has made use of the NASA/IPAC Infrared Science Archive, which is funded by the National Aeronautics and Space Administration and operated by the California Institute of Technology. This work has also made use of Gaia DR3 and public DES DR2 and VHS/VISTA catalogue products. The authors acknowledge the computational resources provided by the Scalable Analytics and Research Virtual Environment (SARVe) facility at the School of Theoretical Physics, IPM.

\facility{Euclid, Gaia, DES, VISTA, IRSA}
\software{Python, PyTorch, NumPy, pandas, Astropy, scikit-learn, Matplotlib}

\section*{Data Availability}

The Euclid Q1 MER and PHZ catalogues used in this study are publicly
available through the Euclid Archive at IRSA
(\href{https://doi.org/10.26131/IRSA602}{MER} and
\href{https://doi.org/10.26131/IRSA604}{PHZ}).
All external catalogues used to construct the reference sample and
cross-match candidates are publicly available, including the
spectroscopic quasar catalogue of \citet{Fu2026}.
The EuclidFM code and data products from this study are available on
\href{https://github.com/AI-Driven-HEP/EuclidFM}{\faGithub}.

\appendix

\section{Data Preparation and EuclidFM Training}
\label{sec:appendix_architecture}

\subsection{Working Catalog and Quality Selection}
\label{sec:appendix_sample}

The parent catalog contains 100,000 Euclid Q1 objects selected randomly
within the three Deep Fields, with approximately equal numbers from each
field: 33,334 from EDF-North and 33,333 each from EDF-South and EDF-Fornax.
We apply the following requirements to the MER detection-quality and
spurious-source flags:
\begin{align}
\texttt{MER\_\_DET\_QUALITY\_FLAG} &\in \{0,2,512\},
\label{eq:appendix_detection_flags}\\
\texttt{MER\_\_SPURIOUS\_FLAG} &= 0.
\label{eq:appendix_spurious_flag}
\end{align}
The allowed nonzero detection-quality values identify sources flagged as
blended (2) or located within an extended-object area (512), respectively
\citep{EuclidMER2026}. These requirements retain 90,275 objects. The
preprocessing script subsequently excludes 17 additional objects specified
by their Euclid object IDs These 17 objects also occur in the Golden Sample used to construct the supervised quasar reference set. They are therefore excluded from the pretraining working catalog to prevent overlap between the pretraining data and the supervised reference sample. This is an additional sample-independence step, not a further photometric-quality cut. This leaves the 90,258-source working catalog used
here. The final field counts are 29,150 in EDF-North, 30,463 in EDF-South,
and 30,645 in EDF-Fornax.

\subsection{Photometric Channels and Availability}
\label{sec:appendix_channels}

The input photometry comprises Euclid measurements and the DECam
\citep{Flaugher2015,DESDR2}, HSC \citep{Aihara2018}, Pan-STARRS
\citep{Chambers2016}, and MegaCam \citep{Ibata2017} measurements included
in the Q1 catalog products. The Euclid channels cover $I_{\rm E}$ from VIS
and $Y_{\rm E}$, $J_{\rm E}$, and $H_{\rm E}$ from NISP, together with
aperture measurements on the combined NISP detection image,
\texttt{NIR\_STACK}. This stack is a composite image rather than an
additional independent passband. The selected external measurements cover
$u$, $g$, $r$, $i$, and $z$, with the available bands depending on the
survey. The measurement-type labels distinguish four aperture sizes
(FWHM1--FWHM4), template fitting (TemplFit), S\'ersic model fitting,
unified photometry (Unif), and VIS PSF fitting. The Unif values are PHZ
fluxes corrected for Galactic extinction and filter shifts.\footnote{See
the \href{https://euclid.esac.esa.int/dr/q1/dpdd/phzdpd/dpcards/phz_phzpfoutputcatalog.html}{Euclid Q1 PHZ data-product description}.}
In the selected layout,
82 flux--uncertainty pairs come from MER columns and 12 Unif pairs from
PHZ columns.

Each channel is identified by a triplet,
\begin{equation}
c_j=(s_j,b_j,t_j),
\label{eq:appendix_channel_identity}
\end{equation}
where $s_j$, $b_j$, and $t_j$ denote the survey, band, and measurement
procedure. The five survey labels, ten band labels (including
\texttt{NIR\_STACK}), and eight procedure labels formally allow
$5\times10\times8=400$ triplets. Many combinations are absent because
the surveys provide different bands and procedures. The preprocessing
identifies paired flux and uncertainty columns and excludes a pair when
either column is entirely zero or missing. This gives $L=94$ channels:
31 Euclid, 28 DECam, 14 HSC, 14 MegaCam, and 7 Pan-STARRS channels.
Several channels measure the same band using related procedures and can
share underlying images. The channel count therefore does not quantify
the number of independent physical observables.

All objects share the same channel layout. Channels are sorted using a
fixed lookup of representative band wavelengths, with ties retaining the
catalog-column order. The survey, band, and procedure identifiers, together
with the original channel index, identify each token even when other
channels are unavailable. For object $i$, channel $j$ is available when
both its flux $F_{ij}$ and reported uncertainty $\sigma_{F,ij}$ are finite.
This describes the saved preprocessing mask; no additional positivity or
signal-to-noise requirement is applied when constructing it. Non-finite
scalars are replaced by zeros for storage, and their pairs are excluded
using the availability mask. Only available, unmasked channels enter the
encoder; padding used to form batches is ignored by attention. Thus,
94 is the size of the shared channel inventory, while the number of
available measurements varies between objects. In the working catalog,
the minimum, maximum, and median numbers of available flux--uncertainty
pairs per object are 29, 70, and 64, respectively.

\subsection{Flux Transformation and Apparent Magnitudes}
\label{sec:appendix_flux_transform}

We express the catalog fluxes and their uncertainties in $\mu$Jy and apply
the inverse hyperbolic sine transformation separately to both quantities:
\begin{equation}
\bm{x}_{ij}=\left[
\asinh\!\left(\frac{F_{ij}}{F_0}\right),\,
\asinh\!\left(\frac{\sigma_{F,ij}}{F_0}\right)
\right],\qquad F_0=1\,\mu{\rm Jy}.
\label{eq:appendix_asinh}
\end{equation}
The transformation is approximately linear near zero and
logarithmic at large absolute values. It therefore compresses the wide
flux range while remaining defined and monotonic for zero and negative
background-subtracted fluxes \citep[cf.][]{Lupton1999}. No fitted scaling
parameters are required. The transformed flux and uncertainty are the two
continuous features of each token; the survey, band, and procedure are
encoded separately as categorical features.

\subsection{Architecture and Pretraining}
\label{sec:appendix_pretraining}

For each source, EuclidFM samples
\begin{equation}
M_i=\min(20,N_{{\rm valid},i})
\end{equation}
available channels without replacement and removes them from the encoder
input, where $N_{{\rm valid},i}$ is the number of available channels.
Unavailable channels are never reconstruction targets. The flux and
uncertainty scalars are separately projected through
$1\!\rightarrow\!64\!\rightarrow\!64$ GELU MLPs. Band, survey, and
measurement-type embeddings have dimensions 8, 8, and 4. Their
concatenation is fused to a 128-dimensional token, and learned positional
embeddings use each channel's \emph{original} sequence index. A learned
\texttt{[CLS]} token is prepended.

The encoder has eight Transformer layers, $\dmodel=128$, eight attention
heads, feed-forward width 512, and dropout 0.1. During pretraining the
\texttt{[CLS]} token is not sent to the reconstruction decoder. Visible
measurement embeddings are restored to their original locations, masked
locations receive a learned mask token, and unavailable locations are
padding-masked. A four-layer decoder predicts transformed flux and
uncertainty through a $128\!\rightarrow\!256\!\rightarrow\!128\!\rightarrow\!2$
head. The reconstruction objective is evaluated only on masked available
positions,
\begin{equation}
\mathcal{L}_{\rm rec}=\frac{1}{\sum_i M_i}
\sum_i\sum_{j\in\mathcal{M}_i}
\frac{\|\widehat{\bm{x}}_{ij}-\bm{x}_{ij}\|_2^2}{2},
\label{eq:loss}
\end{equation}
where $\mathcal{M}_i$ is the set of channels withheld for object $i$.
Figure~\ref{fig:architecture} summarizes the complete data flow. The optional
global-feature branch shown in the schematic is disabled for the results
reported here.

The 90,258-source working catalog is divided into 72,206 training and
18,052 validation objects, an 80:20 split grouped by Euclid object ID.
Pretraining uses AdamW \citep{Loshchilov2019} with learning rate
$3\times10^{-4}$, weight decay 0.01, batch size 64, a configured limit of
1,000 epochs, gradient-norm clipping at 1.0, and seed 42. The learning rate
is halved after ten validation epochs without improvement. A checkpoint
is a saved training state containing the model weights, optimizer and
scheduler states, epoch number, and loss values. We use the checkpoint
with the lowest validation loss, obtained at epoch 656 in the
reported run. After pretraining, we omit the
decoder and extract the encoder's 128-dimensional \texttt{[CLS]} output
using all available measurement tokens. We use t-SNE
\citep{vanDerMaaten2008} only to visualize latent structure; its coordinates
are not classifier inputs or physical distance measures.

\section{High-purity three-class reference set}
\label{sec:platinum}

The supervised reference catalogue is constructed from
spectroscopically confirmed quasars and independently
validated stellar and galaxy candidates. Initial stellar
and galaxy candidates are selected from Euclid PHZ
classification codes 1 and 2, respectively, within the
survey tiles containing the Golden Sample quasars.
For these candidates, we require
\texttt{PHZ\_\_PHZ\_FLAGS}=0,
\texttt{MER\_\_DET\_QUALITY\_FLAG}=0,
\texttt{MER\_\_SPURIOUS\_FLAG}=0,
\texttt{MER\_\_FLAG\_VIS}=0, and
\texttt{MER\_\_FLAG\_NIR\_STACK}=0.
The PHZ classifications are used only for candidate
preselection; the final reference labels are assigned
using independent catalogue evidence.

Stellar candidates are validated using Gaia DR3
astrometry \citep{GaiaDR3}. Euclid Gaia identifiers
are matched directly to Gaia \texttt{source\_id},
requiring a unique Gaia association for each accepted
Euclid source. We require a finite RUWE in the range
$0 \leq \mathrm{RUWE}<1.4$, at least nine visibility
periods, \texttt{astrometric\_params\_solved}
$\in\{31,95\}$, and
\texttt{duplicated\_source}=\texttt{false}.
These criteria select sources with acceptable
astrometric solutions and adequate observational
coverage.

In addition, each accepted star must exhibit either
significant positive parallax or significant proper
motion. The adopted criterion is
\begin{equation}
\frac{\varpi}{\sigma_\varpi}\geq 5
\quad \mathrm{or} \quad
Q_\mu \geq 25,
\end{equation}
where
\begin{equation}
Q_\mu =
\frac{a^2+b^2-2\rho ab}{1-\rho^2},
\end{equation}
with
\begin{equation}
a=\frac{\mu_{\alpha }}{\sigma_{\mu_{\alpha }}},
\qquad
b=\frac{\mu_\delta}{\sigma_{\mu_\delta}},
\qquad
\rho=\texttt{pmra\_pmdec\_corr}.
\end{equation}
Here, $\rho$ denotes the Gaia-reported correlation
coefficient between the two proper-motion components.
The proper-motion statistic accounts for their
measurement covariance, with positive uncertainties
and $|\rho|<1$ required for its evaluation.
The positive-parallax criterion excludes significant
negative-parallax measurements as direct evidence
of stellar parallax.

We further assess the positional consistency of each
Gaia counterpart by linearly propagating its
coordinates using the reported proper motions to
epochs 2024.12 and 2024.77, which bracket the
adopted Euclid Q1 observation interval. The angular
separation from the Euclid position must not exceed
$0.7$ arcsec at either epoch. This is a conservative
linear positional-consistency check, rather than a
full covariance-propagated astrometric cross-match.
A total of 3,500 accepted stars are retained.

Galaxy candidates are independently validated using
DES DR2 \citep{DESDR2} and VHS DR4
\citep{McMahon2013}. Positional cross-matches are
performed within a radius of $0.7$ arcsec.
Ambiguous associations, including multiple
counterparts to an individual Euclid source or
counterparts shared by multiple Euclid candidates,
are rejected.

For DES, we require the extended-source morphology
classification \texttt{ExtClsCoad}=3, corresponding
to the strongest extended-source category used
in our selection. Additionally, at least two of
the DES $g$, $r$, $i$, and $z$ bands must have
integer photometric quality flags between 0 and 3,
inclusive.

An accepted DES counterpart must also have an
unambiguous VHS DR4 association within $0.7$
arcsec. We require \texttt{PriOrSec}=0, selecting
primary VHS detections, together with either
\texttt{Mclass}=1 or
\begin{equation}
p_{\rm Gal}\geq0.8,\qquad
p_{\rm Gal}>p_{\rm Star},\qquad
p_{\rm Gal}>p_{\rm Noise}.
\end{equation}
Thus, both DES and VHS provide required external
evidence for the accepted galaxy reference sample,
rather than VHS being used only as an optional
consistency check.

\begin{table}[!htbp]
\caption{Composition of the final Platinum Sample.}
\label{tab:sample}
\centering
\begin{tabular}{@{}lrl@{}}
\toprule
Class & $N$ & Principal evidence \\
\midrule
Quasar & 2,675 & Euclid Q1 spectroscopy \\
Star & 3,500 & Gaia DR3 astrometry \\
Galaxy & 3,500 & DES DR2 and VHS DR4 \\
\midrule
Total & 9,675 & Unique Euclid object IDs \\
\bottomrule
\end{tabular}
\end{table}

Before final selection, stellar and galaxy candidates
must have valid finite flux--uncertainty pairs with
strictly positive uncertainties in at least three
distinct Euclid bands among VIS, Y, J, and H.
At least one valid Euclid-band measurement must
satisfy a signal-to-noise ratio of five or greater.
Candidates with another accepted candidate or
Golden Sample quasar within 1 arcsec are excluded
to reduce potential source confusion.

After these quality, external-validation, photometric
coverage, and close-pair criteria are applied,
3,500 galaxies are retained. The final stellar
and galaxy selections use fixed target counts
and a reproducible sampling procedure.

The final Platinum Sample comprises 9,675 unique Euclid
sources, including 2,675 spectroscopically confirmed
quasars selected from the Euclid Q1 Golden Sample,
3,500 Gaia DR3-validated stars, and 3,500 galaxies
validated through DES DR2 and VHS DR4.

The quasar subset is obtained after applying the
adopted quality and exclusion criteria to the
spectroscopically confirmed Golden Sample sources.
The resulting three-class sample is used for the
downstream supervised classification analysis.

\newpage
\section{How We Test Similarity in CLS Space}

This appendix explains the tests in Section~\ref{sec:results}.
The t-SNE map is only a visual
aid, all distances are Euclidean distances between the original
128-dimensional CLS vectors.
We first measure a statistic using the catalogue labels or values as
observed. We then keep the vectors, the list of neighbours or comparison
pairs, and the group membership of each source fixed. Within each group,
we randomly reassign the labels or values being tested, without
replacement, and recompute the same statistic. For example, shuffling
point-like labels within a field and a 0.25-mag VIS bin changes which
sources are labelled high but not how many high labels the bin contains.
The shuffled scores describe what we would expect if the assignment of
labels or values to particular CLS vectors were random within the groups.

\subsection{Counting Labels Among Neighbours}
\label{sec:appendix_cls_tests_2}

\paragraph{Point-like probability.}
To construct the randomized baseline, we shuffle the high and low labels within groups defined by the Euclid field and VIS magnitude. This preserves the number of sources in each label category within each group, while removing their original association with positions in the embedding space. The embeddings and nearest-neighbour relationships remain unchanged. After each shuffle, we recalculate the clustering score using the reassigned labels. Importantly, the shuffled labels determine both which sources contribute to the average and which of their neighbours are counted as high-labelled. We repeat this procedure 99 times and take the mean of the resulting scores.

\paragraph{Deblending.}
Here we keep each source's Kron-radius group fixed (largest decile or
the rest) and compare the mean fraction of flagged neighbours between
the two groups. The observed difference is
$100(2.85-1.20)/15=11.0$ percentage points. Shuffling only the deblend
flags within field and magnitude bins preserves the flag rate in each
bin, but not necessarily the rate within each size group. Shuffling
separately by size group within each bin preserves both. Each comparison
uses 999 shuffles. The observed difference is consistent with the second
shuffled baseline: large sources can have more flagged neighbours if
they tend to lie near other large sources and large sources are more
often flagged.

\subsection{Comparing Pairs of Sources}
\label{sec:appendix_cls_tests_mid}

For each source used as a reference, we draw eight pairs of distinct
comparison sources from its group, excluding the reference itself.
Call the reference $i$, the closer member
in CLS space $j$, and the farther member $k$. For catalogue quantity
$x$, the pair agrees if $|x_j-x_i|<|x_k-x_i|$. An agreement scores one,
a disagreement zero, and a tie in either distance half. For VIS
magnitude the group is the field. For each of the 999 shuffles we
keep the same pairs and reassign catalogue values within the original
groups. The reassignment includes the reference source. The shuffled
scores are near 50\% because neither member is then favoured to be
closer in the catalogue quantity.

\paragraph{VIS measurement quality.}
We use the MER VIS flux $F_I$ and its reported uncertainty $\sigma_I$, measured in the same $2\,\mathrm{FWHM}$-diameter aperture used for $I_{\rm E}$ \citep{EuclidMER2026}. The two quantities considered are $\log_{10}(F_I/\sigma_I)$ and $\log_{10}(\sigma_I/\mu\mathrm{Jy})$. We first draw comparison pairs from sources within the same Euclid field, using all 88,576 sources with positive, finite flux and uncertainty. We then repeat the analysis with comparison pairs drawn from sources in the same field and original 0.25-mag $I_{\rm E}$ bin. This leaves 88,535 sources in groups containing at least three members. For the randomized tests, we shuffle the catalogue values within the same groups used to select comparison pairs, while keeping the original magnitude-bin assignments fixed. Finally, we repeat both analyses after excluding sources with $\sigma_I>1\,\mu\mathrm{Jy}$. This leaves 86,550 sources, of which 86,513 belong to field-and-magnitude groups containing at least three members.

\subsection{Fitting in Two Fields and Testing in the Third}
\label{sec:appendix_cls_tests_last}
This is a separate test with no shuffling. We fit a least-squares model
of the form $\widehat{x}=a+\bm{w}^{\mathsf{T}}\bm{z}$, where $\bm{z}$
is the frozen CLS vector. For example, we fit the coefficients using
North and South, then predict VIS magnitude in Fornax without refitting.
We repeat with each field held out in turn and report the Spearman rank
correlation between predicted and measured values in that field. Only
the linear fit holds out a field; the encoder is not retrained for each
split. For the VIS measurement-quality tests, we use the same log S/N
and log uncertainty targets as in the pair comparisons and fit the
models on the full and restricted samples separately. For log flux
uncertainty we also compare a magnitude-only cubic spline with the
same spline basis plus the CLS coordinates. Both models and the spline
basis (eight knots at training-sample magnitude quantiles) are fitted
using only the two training fields. Their root-mean-square errors
are compared on the same held-out sources.

\bibliography{references}
\bibliographystyle{aasjournal}

\end{document}